\documentclass[journal=jpcbfk,manuscript=article]{achemso}

\usepackage[version=3]{mhchem}
\usepackage[T1]{fontenc}
\usepackage{amsmath}
\usepackage{amssymb}
\usepackage{color}
\usepackage{graphicx}
\usepackage{dcolumn}
\usepackage{bm}
\usepackage{hyperref}
\usepackage[b]{esvect}

\newcommand{\rp}{\mathbf{r}}
\newcommand{\np}{\mathbf{\hat{n}}}
\newcommand{\bp}{\mathbf{b}}

\DeclareMathOperator*{\argmax}{arg\,max}

\author{Sucheol Shin}
\author{Adam P. Willard}
\affiliation{Department of Chemistry, Massachusetts Institute of Technology, Cambridge, Massachusetts 02139, Unites States}
\email{awillard@mit.edu}

\title[An \textsf{achemso} demo]
  {Water's interfacial hydrogen bonding structure reveals the effective strength of surface-water interactions}

\begin{document}

\begin{abstract}
The interactions of a hydrophilic surface with water can significantly influence the characteristics of the liquid water interface. 
In this manuscript, we explore this influence by studying the molecular structure of liquid water at a disordered surface with tunable surface-water interactions. 
We combine all-atom molecular dynamics simulations with a mean field model of interfacial hydrogen bonding to analyze the effect of surface-water interactions on the structural and energetic properties of the liquid water interface. 
We find that the molecular structure of water at a weakly interacting (\emph{i.e.}, hydrophobic) surface is resistant to change unless the strength of surface-water interactions are above a certain threshold. 
We find that below this threshold water's interfacial structure is homogeneous and insensitive to the details of the disordered surface, however, above this threshold water's interfacial structure is heterogeneous. 
Despite this heterogeneity, we demonstrate that the equilibrium distribution of molecular orientations can be used to quantify the energetic component of the surface-water interactions that contribute specifically to modifying the interfacial hydrogen bonding network.
We identify this specific energetic component as a new measure of hydrophilicity, which we refer to as the intrinsic hydropathy. 
\end{abstract}

\section{Introduction}

In the vicinity of an extended hydrophilic surface, aqueous properties such as molecular mobility, solute solubility, and chemical reactivity can differ significantly from their bulk values \cite{Eisenthal1996, Benjamin1996, Ge2006, Brindza2009, Patel2011a, Mante2014, Bjorneholm2016a, Wen2016}.
These differences reflect the characteristics of water's interfacial hydrogen bonding network and how these characteristics are influenced by the presence of surface-water interactions.
The effect of these interactions are difficult to predict due to the collective structure of the aqueous interfacial hydrogen bonding network. 
Understanding the influence of surface-water interactions on this interfacial hydrogen bonding network is therefore fundamental to the study of hydrophilic solvation.

In this manuscript we investigate the response of the hydrogen bonding network to changes in the strength of surface-water interactions.
We present the results of molecular dynamics simulations of the interface between liquid water and a model surface with tunable hydrophilicity.
We utilize a rigid model surface that is molecularly disordered and includes polarized (\emph{i.e.}, hydrogen bond-like) interactions that have heterogeneous orientations.  
We examine the structure of the interfacial hydrogen bonding network and how it varies when the strength of the surface-water interactions are changed. 
We find that hydrophilic surfaces, with surface-water interactions that are similar in strength to typical aqueous hydrogen bonds, give rise to interfacial molecular structure that is spatially heterogeneous. 
As we demonstrate, this heterogeneous structure includes some regions with interfacial molecular structure that is only weakly perturbed from that observed at an ideal hydrophobic surface. 
As we highlight, the persistence of this weakly perturbed, hydrophobic-like interfacial molecular structure may explain the ubiquity of hydrophobic effects in aqueous solvation.

The molecular structure of a liquid water interface is determined primarily by water's strong tendency to engage in tetrahedrally coordinated hydrogen bonding \cite{Shin2017}.
In the bulk liquid this tendency leads to the formation of a disordered tetrahedral hydrogen bonding network. 
The characteristics of this network determine many of water's physical properties, such as its density, heat capacity, and viscosity \cite{Stillinger1980}. 
The individual hydrogen bonds that comprise this network are very energetically favorable, so any given bond within the bulk liquid is broken only fleetingly.\cite{Eaves2005b}
At an interface, however, geometric constraints make it impossible to simultaneously satisfy all available hydrogen bonds. 
Molecules at the interface thus reorganize to mitigate the loss of hydrogen bonds resulting in an interfacial hydrogen bonding network that is anisotropic and distorted relative to that of the bulk liquid \cite{Geissler2013}.
Aqueous properties that depend on this network structure, such as small molecule solvation\cite{Turner1994,Vanzi1998} and proton transport\cite{Geissler2001,Kattirtzi2017}, thus vary in the vicinity of a liquid water interface. 

Notably, the characteristics of water's interfacial molecular structure can be altered by the presence of external interactions, such as those that arise at a hydrophilic surface. 
Many previous studies have been aimed at revealing the microscopic properties of water at hydrophilic surfaces. 
Experimental efforts, such as those based on sum-frequency generation spectroscopy\cite{Shultz2000, Richmond2002, Shen2006}, terahertz absorption spectroscopy\cite{Ebbinghaus2007,Heyden2008}, and dynamic nuclear polarization\cite{McCarney2008,Armstrong2009}, have uncovered important details about the microscopic structure and dynamics of the liquid water interface.
These efforts have revealed that strong surface-water interactions can significantly reduce the mobility of interfacial water molecules and modify aqueous hydrogen bonding energetics \cite{Heugen2006,Armstrong2011,Wen2016}.
Theoretical efforts, such as those based on first-principles calculations\cite{Heyden2010,Gaigeot2012}, classical molecular dynamics (MD) simulations \cite{Giovambattista2007b,ContiNibali2014}, and continuum\cite{Schlaich2016a} or coarse-grained modeling\cite{Marrink2007} have provided fundamental physical insight into the role of hydrogen bonding in determining microscopic interfacial structure and have been vital to the interpretation of many experimental results.
surface-water
Here we build upon these previous studies with a model system that is designed to allow for systematic variations in surface-water interactions.
Our approach is unique because we use the collective structure of the interfacial hydrogen bonding network as an order parameter for quantifying water's interfacial molecular structure. 
We use this order parameter to resolve the spatial dependence of water's response to disordered but strongly interacting surfaces.
We complement this approach by using a mean field model of interfacial hydrogen bonding to isolate the energetic component of these surface-water interactions that contribute specifically to reorganizing water's interfacial hydrogen bonding network. 
We then propose that this particular energetic component represents an intuitive measure of surface hydropathy.

Details about our model system and the methods we use to analyze and characterize water's interfacial molecular structure are described in the following section. 
Then, in the section entitled ``The effect of surface polarity on water's interfacial molecular structure,'' we present results and discuss how variations in surface-water interactions affect water's interfacial molecular structure.
In the section entitled ``Quantifying surface hydropathy from water's interfacial molecular structure,'' we describe a mean field model of interfacial hydrogen bonding and show how this model can be applied to quantify surface hydropathy. 

\section{Model and Methods}
\subsection{A disordered model surface with tunable hydrophilicity}
\begin{figure}[h!]
\centering
\includegraphics[width = 3.4 in]{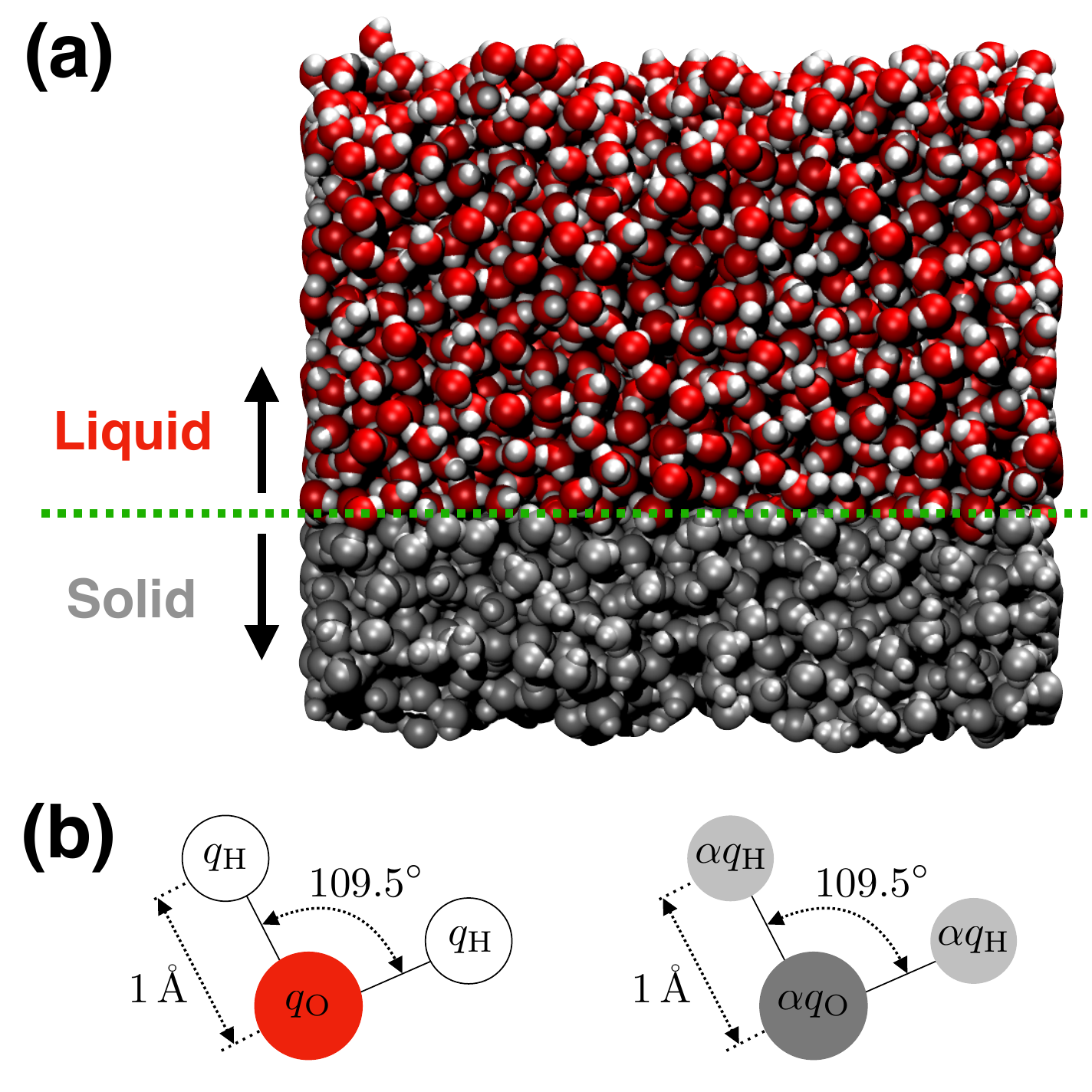}
\caption{(a) A simulation snapshot of the model system. The disordered model surface is represented by grey-colored water molecules which are immobilized during the simulation. 
The liquid is represented by red and white-colored molecules. 
The green dotted line indicates the approximate location of the liquid-surface interface. 
(b) A schematic illustration highlighting the difference between standard SPC/E water molecules, which are described as three point charges arranged with a specific relative geometry and embedded within a Lennard-Jones potential, and the surface molecules, which are modeled identically except that the point charges are scaled by a factor of $\alpha$.
}
\label{fig:1}
\end{figure}

Our model surface is constructed from an immobilized slab of bulk liquid water with variable partial charges that can be used to tune surface hydrophilicity.
This model surface has been previously used to investigate the influence of surface-water interactions on interfacial density fluctuations.\cite{Patel2010,Willard2014}
As illustrated in Fig.~\ref{fig:1}(a), our surface is constructed based on an equilibrium configuration of a slab of liquid water, spanning the periodic boundaries of the $x\text{-}y$ plane.
We define the surface as the set of water molecules whose oxygen atoms lie on one side of a plane perpendicular to the $z$-axis that cuts through the liquid water slab.
The position of this plane is located with a value of $z$ that is far enough from either interface as to characterize properties of the bulk liquid.
Water molecules belonging to the surface are thus immobilized, fixed in a single configuration that is representative of the equilibrium bulk liquid. 

Water molecules belonging to the model surface interact with a mobile population of ordinary water molecules via the standard water-water interaction potential.
For the results presented below we utilize the SPC/E model of water,\cite{Berendsen1987} however, this surface construction could be applied to any classical atomistic model of water.
We tune the surface-water interactions by scaling the partial charges of the immobilized surface molecules by a factor of $\alpha$, thereby scaling the polar hydrogen bonding interactions of surface molecules with those of the liquid.
The charges on the surface oxygens and hydrogens are therefore given by $q^{(\mathrm{surf})}_\mathrm{O}=\alpha q_\mathrm{O}$ and $q^{(\mathrm{surf})}_\mathrm{H}=\alpha q_\mathrm{H}$, where $q_\mathrm{O}$ and $q_\mathrm{H}$ are the partial charges of the SPC/E model. 
Scaling the surface charges in this way preserves the charge neutrality of the surface.

To prepare the model surfaces that are used in the results described below we first equilibrated slab 3564 water molecules in a periodically replicated simulation cell with dimensions $5 \times 5 \times 12 \text{ nm}^3$ at 298 K.
The simulation cell is longer in the $z$-dimension so that the liquid water spontaneously forms a slab that is approximately $4.5 \text{ nm}$ in thickness that contains two separated water-vapor interfaces.
For a given configuration of the slab we defined the surface by drawing a horizontal plane through the liquid slab at a vertical position approximately $1.2 \text{ nm}$ from the lower water-vapor interface.
The resulting model surfaces are thus approximately 1.2 nm thick and contain about 1000 immobilized water molecules.

We used the above procedure to generate five different model surfaces based on independent equilibrium slab configurations. 
For each surface and each value of $\alpha$, the dynamic population of non-surface (\emph{i.e.}, liquid) water molecules were allowed to equilibrate in the presence of the surface for 0.1 ns at 298 K prior to gathering statistics. 
All simulations were performed in the NVT ensembles with a Langevin thermostat with the LAMMPS simulation package \cite{Plimpton1995}.
Details about the simulation setup can be found in the Supporting Information (SI). 

\subsection{A method for quantifying the molecular structure of a liquid water interface \label{sec:methods}}

We characterize the molecular structure of the liquid water interface by analyzing the orientational statistics of interfacial water molecules.
To do this we utilize a structural order parameter, $\delta \lambda_\mathrm{phob}$, that quantifies how these orientational statistics differ from those that arise at an ideal hydrophobic surface.
As described in Ref.~\citenum{Shin2018}, this order parameter is capable of distinguishing hydrophobic and hydrophilic surfaces based only on the effect of these surfaces on aqueous interfacial molecular structure.
Furthermore, $\delta \lambda_\mathrm{phob}$ can be formulated as a local order parameter to generate spatially resolved maps of water's interfacial molecular structure. 
We summarize the formulation of $\delta \lambda_\mathrm{phob}$ below. 
A more complete description of this order parameter can be found in Ref.~\citenum{Shin2018}.

To calculate $\delta \lambda_\mathrm{phob}$ we first sample the orientational configurations of interfacial water molecules.
We denote the orientational configuration of a water molecule in terms of the three-dimensional vector, $\vec{\,\xi\,}=(\cos \theta_1,\cos \theta_2, a)$, where $\theta_1$ and $\theta_2$ specify the angle made between each OH bond vector and the local surface normal and $a$ specifies the distance of the water molecule from the nearest position of the instantaneous liquid water interface\cite{Willard2010}.
We define the position of the instantaneous liquid water interface following the procedure described in Ref.~\citenum{Willard2010}.
For any given orientational configuration we can compute the quantity
\begin{equation}
f\big(\vec{\,\xi\,}\big\vert\text{phob}\big) = -\ln\left[\frac{P\big(\vec{\,\xi\,}\big|\text{phob}\big)}{P\big(\vec{\,\xi\,}\big|\text{iso}\big)}\right]~,
\label{eq:FE}
\end{equation}
where $P\big(\vec{\,\xi\,}\big\vert\text{phob}\big)$ denotes the pre-tabulated probability to observe the specific molecular orientation, $\vec{\,\xi\,}$, at an ideal hydrophobic interface and $P\big(\vec{\,\xi\,}\big\vert\text{iso}\big)$ denote corresponding probability for the case when molecular orientations are distributed isotropically (\emph{e.g.}, within the bulk liquid). 
For a given surface, the quantity $\delta \lambda_\mathrm{phob}$ simply reflects the average value of $f\big(\vec{\,\xi\,} \big\vert \text{phob}\big)$ computed for water molecules at the interface.

Specifically, for a particular location, $\rp_\mathrm{surf}$, along the plane of the liquid-surface interface,
\begin{equation}
\delta \lambda_\mathrm{phob}(\rp_\mathrm{surf}) = \lambda_\mathrm{phob}(\rp_\mathrm{surf}) - \langle \lambda_\mathrm{phob} \rangle_0,
\end{equation}
where,
\begin{equation}
\lambda_\text{phob}(\rp_\text{surf}) = \frac{1}{N_\mathrm{\tau}}\sum_{t=0}^{\tau} f\big(\vec{\,\xi\,}\!(\rp_\text{surf},t)\big|\text{phob}\big)~,
\label{eq:qphob}
\end{equation}
and $\langle \lambda_\mathrm{phob} \rangle_0$ represents the average value of $\lambda_\mathrm{phob}$ computed over an ideal hydrophobic surface.
In Eq.~\ref{eq:qphob}, $\vec{\,\xi\,}\!(\rp_\text{surf},t)$ specifies the orientational configuration of the water molecule that is nearest to the position $\rp_\mathrm{surf}$ at time $t$, the summation is taken over a discrete set of $N_\tau$ simulation snapshots  sampled along the time interval $\tau$.
Here we sample simulation snapshots separated by 100 fs along a 1 ns trajectory (\emph{i.e.}, $\tau=1$ ns and $N_\tau=10000$). 

By definition, $\delta \lambda_\mathrm{phob}\approx 0$ when water's interfacial molecular structure is similar to that of a hydrophobic surface. 
Surfaces that interact strongly with water molecules cause interfacial molecular structure to deviate from that of the hydrophobic reference system, which typically results in positive values for $\delta \lambda_\mathrm{phob}$. 
By computing $\delta \lambda_\mathrm{phob}(\rp_\mathrm{surf})$ locally, we can identify the spatial profile of interfacial distortions that arise due to water's interactions with a heterogeneous surface.

\section{The effect of surface polarity on water's interfacial molecular structure}
We evaluate the effect of surface-water interactions on water's interfacial molecular structure by analyzing simulations carried out using surfaces with different values of $\alpha$. 
We consider values of $\alpha$ ranging from $\alpha=0$, as an example of a disordered hydrophobic surface, to $\alpha=1$, as an example of an ideal (\emph{i.e.}, water-like) hydrophilic surface. 
For each value of $\alpha$ we have considered five independently generated surface configurations.
For each surface configuration we have performed a 1 ns equilibrium simulation. 
We have analyzed each simulation by computing the value of $\delta \lambda_\mathrm{phob}(\rp_\mathrm{surf})$ on a square lattice with lattice spacing equal to $0.5 \text{ \AA}$ along the plane of the liquid-surface interface. 
By definition, variations in $\delta \lambda_\mathrm{phob}$ thus indicate changes in the molecular structure of the liquid water interface. 

To establish an intuitive framework for interpreting variations in interfacial molecular structure we compare values of $\delta \lambda_\mathrm{phob}(\rp_\mathrm{surf})$ to the quantity, $\Delta \mu_\mathrm{ex}(\rp_\mathrm{surf})$, which denotes the change in excess chemical potential of a hard-sphere solute of radius $2.5 \text{ \AA}$ when the solute is brought from the bulk liquid to a position where it contacts the surface at $\rp_\mathrm{surf}$.
$\Delta \mu_\mathrm{ex}(\rp_\mathrm{surf})$ can be used to identify regions of a rigid hydrated surface that are either hydrophobic or hydrophilic\cite{Godawat2009,Patel2014}.
Here, we identify positions of the surface with $\Delta \mu_\mathrm{ex} < -k_\mathrm{B} T$ as being hydrophobic and those with $\Delta \mu_\mathrm{ex}> k_\mathrm{B}T$ as being hydrophilic. 
By comparing the statistics of $\delta \lambda_\mathrm{phob}$ at hydrophobic and hydrophilic surface sites (see SI for more details), we have thus identified the range of $\delta \lambda_\mathrm{phob}$ that correspond to either hydrophobic or hydrophilic interfacial molecular structure.
In particular, we have found that values of $-0.1 \leq \delta \lambda_\mathrm{phob} \leq 0.1$ are indicative of typical hydrophobic molecular structure and values of $\vert \delta \lambda_\mathrm{phob} \vert > 0.1$ are indicative of hydrophilic interfacial molecular structure. 

To analyze our results, we first consider the effect of $\alpha$ on $\langle\delta \lambda_\mathrm{phob}\rangle$, the value of our order parameter averaged over all surface positions and surface realizations. 
As illustrated in Fig. 2, $\langle \delta \lambda_\mathrm{phob} \rangle \approx 0$ for surfaces with $\alpha=0$, indicating that apolar uncharged surfaces give rise to interfacial molecular structure that is characteristically hydrophobic. 
We observe that water's interfacial molecular structure depends weakly on $\alpha$ when $0 \leq \alpha \lesssim 0.4$, suggesting that there is a threshold in surface polarity that must be overcome in order to affect significant change in water's interfacial molecular structure. 
Beyond this threshold, $\langle \delta \lambda_\mathrm{phob} \rangle$ increases steadily and takes on values associated with hydrophilic interfacial structure (\emph{i.e.}, $\langle \delta \lambda_\mathrm{phob} \rangle > 0.1$) when $\alpha \gtrsim 0.6$.

\begin{figure}[h]
\centering
\includegraphics[width = 3.4 in]{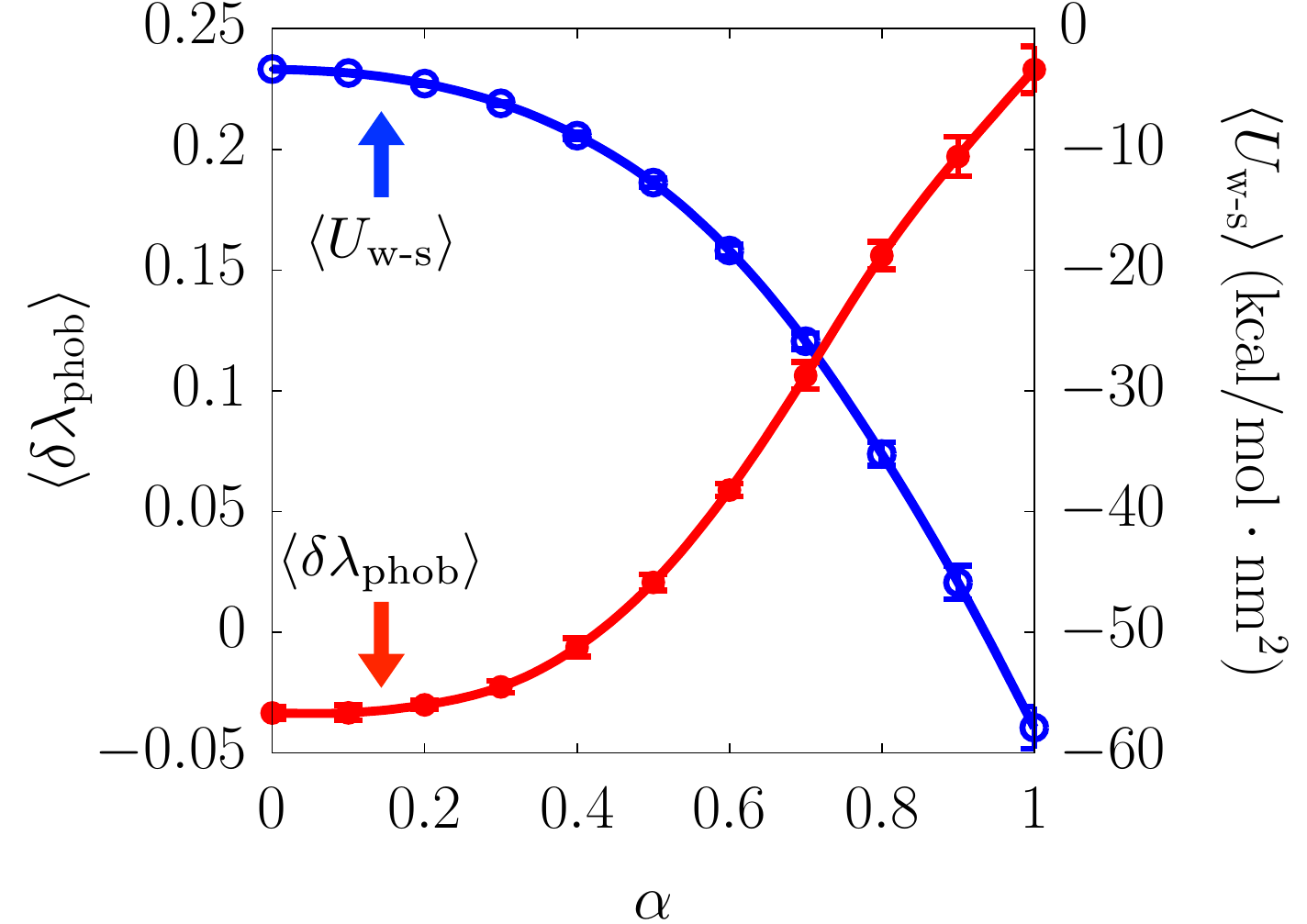}
\caption{A plot of the $\alpha$ dependence of $\langle \delta \lambda_\mathrm{phob} \rangle$ (red line plotted against the left vertical axis) and $\langle U_\text{w-s}\rangle$ (blue line plotted against the right vertical axis).}
\label{fig:2}
\end{figure}

To better understand the effect of $\alpha$ on interfacial molecular structure we have computed the average surface-water interaction energy,
\begin{equation}
\left\langle U_\text{w-s}\right\rangle = \left\langle \sum_{i \in \text{surf}} \sum_{j \in \text{liq}} u_{ij} \right\rangle ~,
\label{eq:u_ws}
\end{equation}
where the angle brackets represent an equilibrium average, the first summation is taken over all frozen surface molecules, the second summation is taken over all molecules in the liquid, and $u_{ij}$ represents the pair potential for interactions between surface species and molecules within the liquid. 
We observe that the dependence of $\langle U_\text{w-s} \rangle$ on $\alpha$ is complementary to that of  $\langle \delta \lambda_\mathrm{phob} \rangle$. 
Specifically, $\langle U_\text{w-s} \rangle$ varies nonlinearly with $\alpha$, slowly for $0\leq \alpha \lesssim 0.4$ and more rapidly for $\alpha \gtrsim 0.4$. 
This complementarity shows that the changes in interfacial molecular structure that are indicated by increases in $\langle \delta \lambda_\mathrm{phob} \rangle$ when $\alpha > 0.4$ are enabled by the formation of more favorable surface-water interactions.
Moreover, the reluctance of $\langle \delta \lambda_\mathrm{phob} \rangle$ to change when the value of $\alpha$ is small illustrates that water's interfacial hydrogen bonding network is determined by a competition between the strength of surface-water and water-water interactions. 
These results thus reveal the strength of favorable surface-water interactions (in terms of $\alpha$) that are required to offset the free energy costs associated with the reorganization of interfacial hydrogen bonding. 

The values of $\langle \delta \lambda_\mathrm{phob} \rangle$ that are plotted in Fig. 2 represent a spatial average over heterogeneous surfaces. 
To understand the effects of surface heterogeneity on local interfacial molecular structure we analyze the statistics of $\delta \lambda_\mathrm{phob}(\rp_\mathrm{surf})$ computed locally at various positions along the liquid-surface interface. 
We characterize the statistics of this local interfacial molecular structure in terms of $P(\delta \lambda_\mathrm{phob})$, the probability to observe a given value of $\delta \lambda_\mathrm{phob}$ at a specfic point along the surface. 
Plots of $P(\delta \lambda_\mathrm{phob})$ computed for surfaces with different values of $\alpha$ are shown in Fig. 3.
We observe that when $\alpha=0$, $P(\delta \lambda_\mathrm{phob})$ is approximately Gaussian with a narrow width centered at $\delta \lambda_\mathrm{phob}\approx0$. 
For larger values of $\alpha$, however, $P(\delta \lambda_\mathrm{phob})$ has pronounced non-Gaussian tails at large values of $\delta \lambda_\mathrm{phob}$. 
Unlike the peak behavior of $P(\delta \lambda_\mathrm{phob})$, which exhibits a small shift with increasing $\alpha$, the large-$\delta \lambda_\mathrm{phob}$ tails are extremely sensitive to changes in $\alpha$. 
These tails indicate that when $\alpha \ge 0.4$, the probability to observe regions with highly distorted interfacial molecular structure is many orders of magnitude larger than would be expected based on Gaussian statistics. 

\begin{figure}[h]
\centering
\includegraphics[width = 3.4 in]{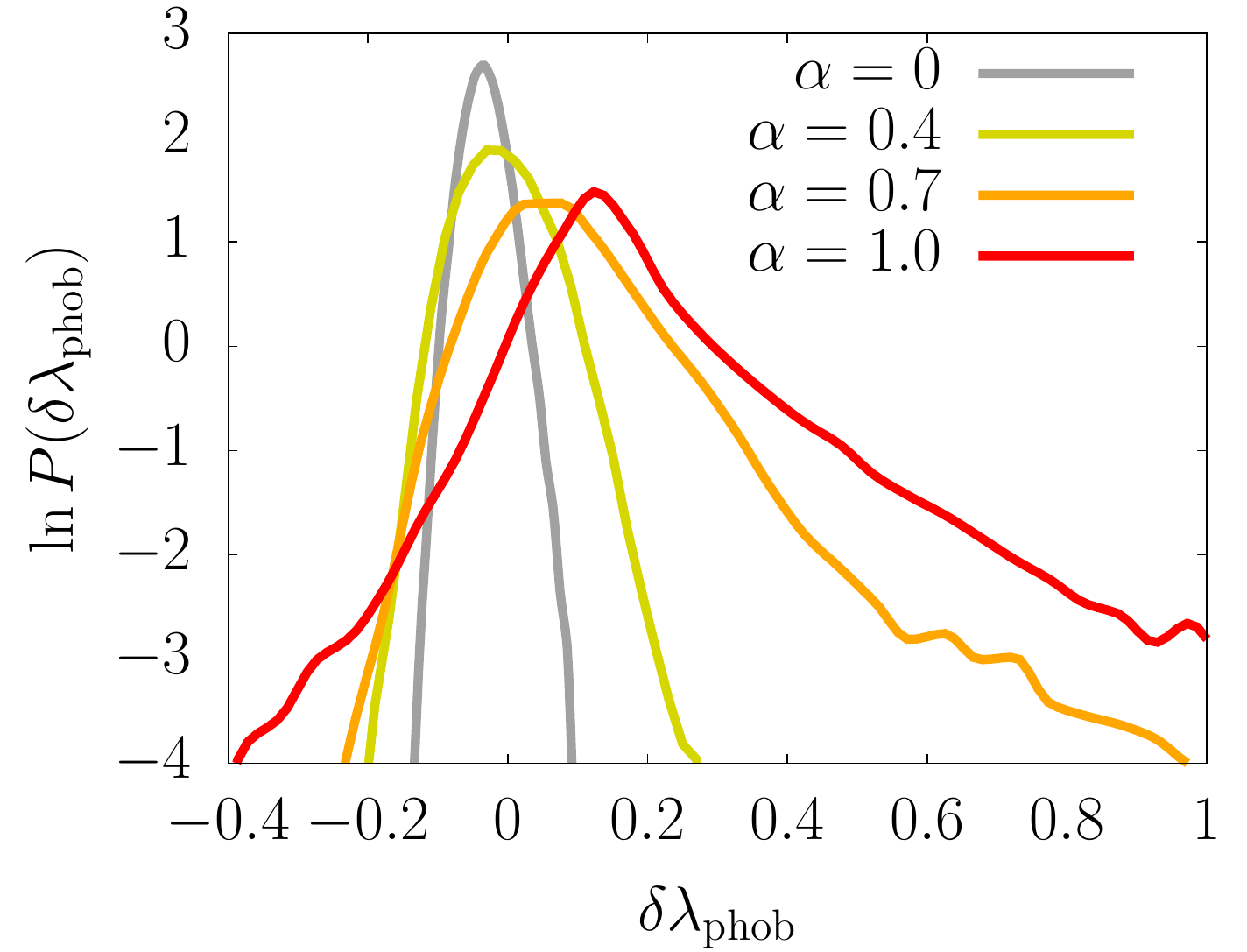}
\caption{A plot of the probability distribution for $\delta \lambda_\mathrm{phob}$ computed for surfaces with different values of $\alpha$. These distributions exhibit mean behavior that shifts systematically with $\alpha$ and the appearance of pronounced non-Gaussian tails at larger values of $\alpha$.}
\label{fig:3}
\end{figure}

The behavior of $P(\delta \lambda_\mathrm{phob})$ indicates that surfaces with $\alpha \geq 0.4$ give rise to aqueous interfacial molecular structure that is heterogeneous. 
To understand the spatial distribution of this heterogeneous interfacial molecular structure we plot $\delta \lambda_\mathrm{phob}(\rp_\text{surf})$ computed for a single fixed surface configuration with different values of $\alpha$. 
The series of panels in Fig. 4 illustrate how $\alpha$ affects the spatial variations in water's interfacial molecular structure. 
When $\alpha$ is small, the structure of the aqueous interface is homogeneous with $\delta \lambda_\mathrm{phob} \approx 0$. 
The spatial distribution of $\delta \lambda_\mathrm{phob}$ is similar for surfaces with $0 \le \alpha \le 0.4$, consistent with the results presented in Figs. 2 and 3. 
When $\alpha > 0.4$, however, we observe the appearance of localized, approximately water-sized domains that have larger values of $\delta \lambda_\mathrm{phob}$. 
These domains correspond to the fat tails in $P(\delta \lambda_\mathrm{phob})$ that are plotted in Fig. 3. 

\begin{figure}[h]
\centering
\includegraphics[width = 6.8in]{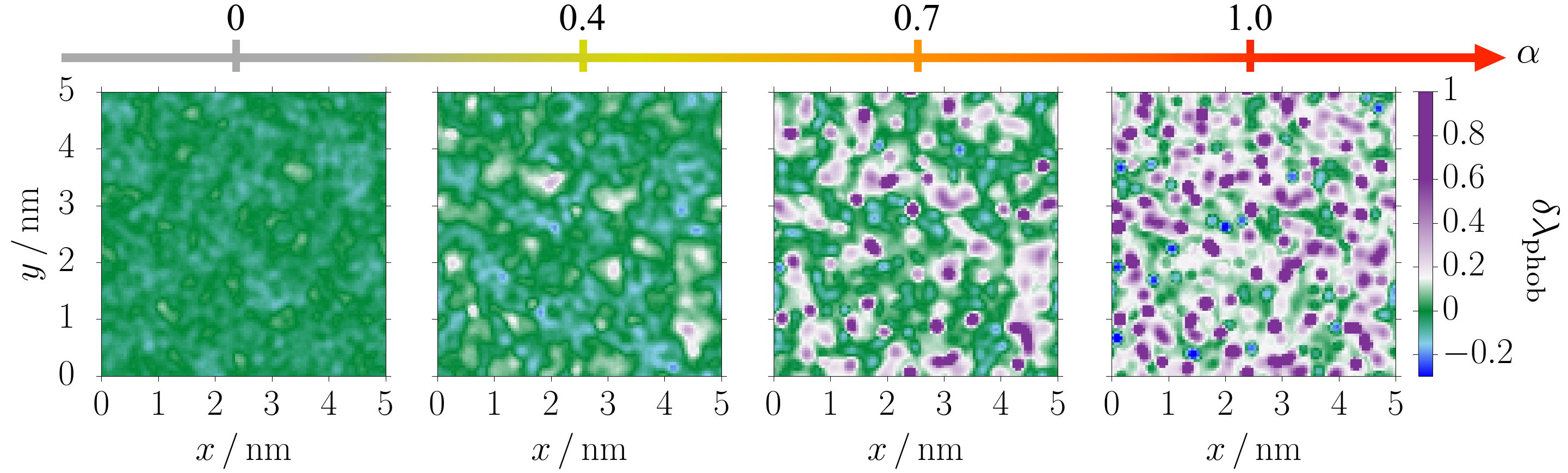}
\caption{
Spatial maps of $\delta \lambda_\mathrm{phob}$ computed for points distributed along the plane of the liquid-surface interface. 
For each value of $\alpha$, the plotted values of $\delta \lambda_\mathrm{phob}$, as indicated by shading, have been averaged over a 1 ns trajectory.
The molecular configuration of the surface is identical for each of the panels.
The color bar is designed to distinguish between regions with interfacial molecular structure that is indicative of hydrophobic surfaces (green shaded regions) and hydrophilic surfaces (purple and blue shaded regions).}
\label{fig:4}
\end{figure}

Notably, even the most hydrophilic surface that we considered (\emph{i.e.}, the $\alpha=1$ surface) includes many regions with interfacial molecular structure that is hydrophobic-like.
In fact, the signatures of this hydrophobic interfacial structure are evident in over 25\% of the interfacial area of the $\alpha=1$ surface.
These regions are associated with the peak behavior of $P(\delta \lambda_\mathrm{phob})$, which as Fig.~\ref{fig:3} illustrates, originates directly from a systematic $\alpha$-induced shift in the $\alpha=0$ distribution.
This observation indicates that the specific molecular structure that is adopted by water at a hydrophobic interface resides in a basin of thermodynamic stability that is robust to moderate surface-induced perturbations.
We attribute the stability of this hydrophobic interfacial molecular structure to the strong influence of the bulk liquid hydrogen bond network on the orientations of interfacial water molecules. 
A surface must overcome this influence in order to cause significant changes in water's interfacial molecular structure. 

The tendency for liquid water to adopt hydrophobic-like interfacial molecular structure, even at hydrophilic surfaces, highlights the importance of this particular interfacial hydrogen bonding arrangement in aqueous solvation.
For instance, this particular interfacial structure determines the thermodynamic driving forces that underlie the hydrophobic effect \cite{Chandler2005}.
Thus, in order to mitigate hydrophobic effects a surface must include a high density of surface sites whose interactions with water molecules are sufficiently strong as to overcome the hydrogen bonding interactions imposed by the adjacent bulk liquid.

Unlike many hydrated surfaces and large solutes, the surfaces we have considered are completely rigid. 
Without this rigidity, the spatial heterogeneity exhibited in Fig. 4 would be absent. 
For dynamic surfaces we expect that heterogeneity in the distribution of $\delta \lambda_\mathrm{phob}$, such as indicated by the fat tails in Fig.~\ref{fig:3}, would still be evident.
However, the presence of spatial heterogeneity, such as illustrated in Fig.~\ref{fig:4}, would be limited to timescales characteristic of surface dynamics.
If the time scale for surface reorganization is similar to that of interfacial water molecules, then evidence of spatial heterogeneity would vanish.
However, for most extended hydrated surfaces, such as those of proteins or other biological macromolecules, surface reorganization is coupled to conformational dynamics and is therefore slow relative to typical solvent dynamics.

\section{Quantifying Surface Hydropathy from Water's Interfacial Molecular Structure}

In this section we introduce the concept of intrinsic hydropathy (IH), which describes the extent to which a hydrated surface alters the intrinsic molecular structure of the liquid water interface.
This quantity is determined by a competition between the constraints imposed on interfacial water molecules by surface-water interactions and those imposed by the collective hydrogen bonding network of the surrounding liquid. 
The outcome of this competition depends specifically on the subset of surface-water interactions that affect the orientational preferences of interfacial water molecules. 
Unfortunately, it is not straightforward to separate these specific interactions from the total set of surface-water interactions, so quantifying their effective strength is challenging.
Here we address this challenge by considering the statistical mechanics of interfacial hydrogen bonding. 

We utilize a mean field model of aqueous interfacial hydrogen bonding at a uniform surface that interacts with water through hydrogen bond-like interactions of tunable strength.
This model surface has a well-defined IH value, which is simply given by the energy of a surface-water hydrogen bond.
By tuning this energy we can determine how water's interfacial molecular structure depends on the value of the IH.
We then exploit this dependence to assign IH values to surfaces based solely on their influence on water's interfacial molecular structure.

In the following subsection we describe the mean field model of aqueous interfacial hydrogen bonding.
Then, we present the application of this model to quantifying the IH of the disordered molecular surfaces described in the previous sections.

\subsection{A Mean-Field Model of Interfacial Hydrogen Bonding at an Interacting Surface}

Here we describe a theoretical model for computing the orientational distribution function of molecules at the interface between liquid water and an interacting surface.
This model is an extension of a similar theoretical framework, introduced in Ref.~\citenum{Shin2017}, for computing interfacial molecular structure at the liquid water-vapor interface.
In Ref.~\citenum{Shin2017}, we show that this model framework can accurately reproduce the primary features of the molecular structure of the water-vapor interface, and here we apply it to describe the water-surface interface. 
We specify interfacial molecular structure in terms of the orientational distribution function for water molecules at various distances from the liquid water interface.
Within this model framework, this distribution function is determined based on the orientational preferences of an individual probe molecule interacting with the average density field of the interfacial environment via an empirical hydrogen bonding potential. 

\begin{figure}[h!]
\centering
\includegraphics[width = 3.4 in]{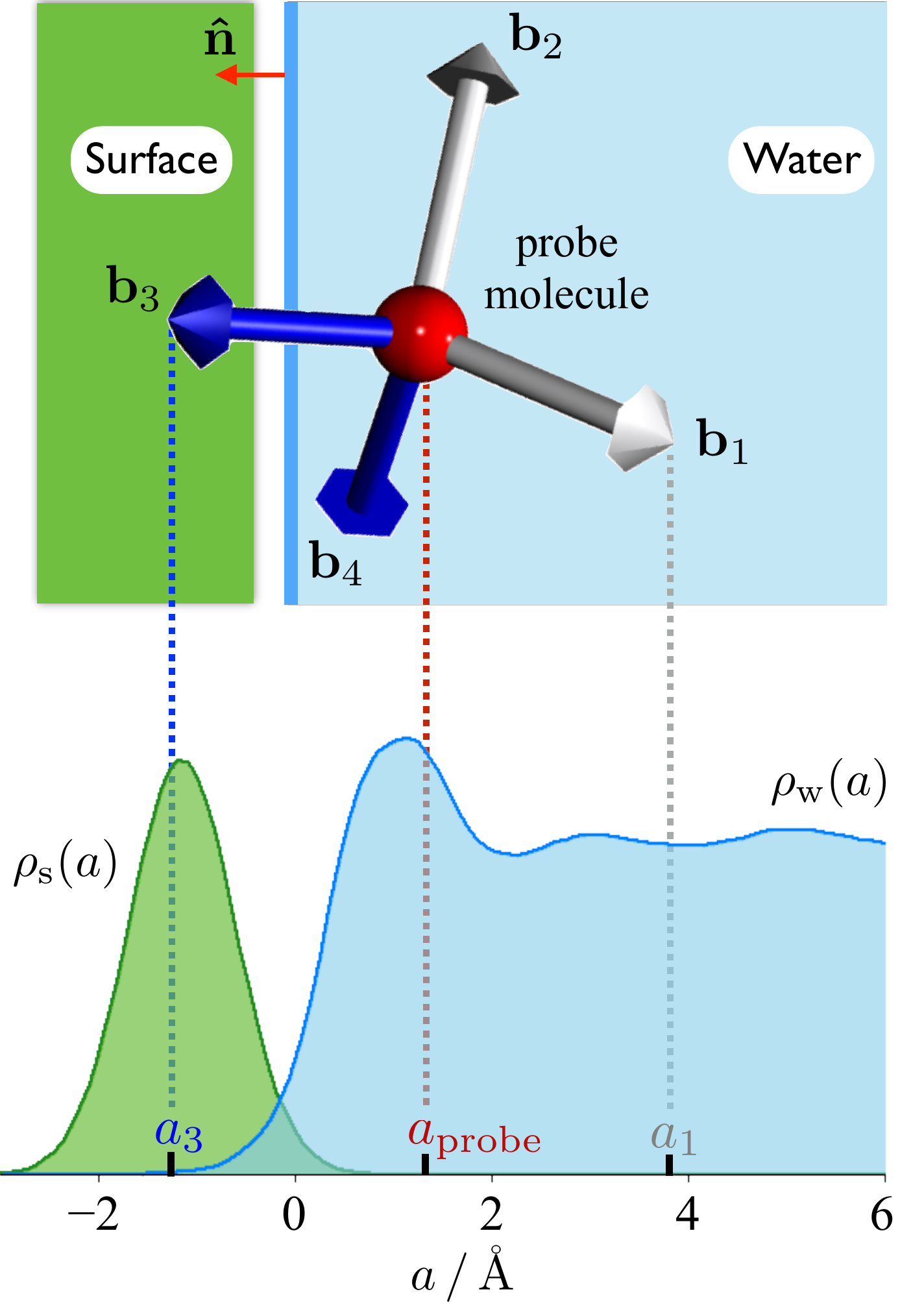}
\caption{Schematic depiction of the mean-field model showing a probe molecule with tetrahedrally coordinated bond vectors (white for donor, blue for acceptor) at a distance $a_\text{probe}$ from the position of the instantaneous interface (solid blue line).
The probe molecule within the liquid phase (blue shaded region) can have hydrogen bonds to either neighboring solvent molecules or nearby solute (green shaded region) through the bond vectors. 
A plot of the interfacial density profile of liquid water, $\rho_\text{w}(a)$, obtained from the MD simulation with SPC/E water, is shown along with that of an empirical solute density profile, $\rho_\text{s}(a)$, where gray and blue dotted lines indicate the termination points of bond vectors $\bp_1$ and $\bp_3$ respectively. }
\label{fig:5}
\end{figure}

As illustrated in Fig.~\ref{fig:5}, the model includes a single probe molecule located at distance $a$ from the position of a planar liquid interface. 
The interfacial environment is described with a density field that is anisotropic in the direction perpendicular to the interface but uniform in the directions parallel to the interface. 
The interfacial density field is composed of two separate elements: a water density, $\rho_\mathrm{w}(a)$, that is computed form atomistic simulation and a surface density, $\rho_\mathrm{s}(a)$, that represents the distribution of interacting sites on the extended model surface.
As depicted in Fig.~\ref{fig:5}, we approximate this distribution as being Gaussian with characteristics that reflect the molecular roughness of a given hydrated surface.

The probe molecule is described as a point particle with four tetrahedrally coordinated hydrogen bond vectors, denoted $\bp_1$, $\bp_2$, $\bp_3$, and $\bp_4$ (see Fig.~\ref{fig:5}).
The length of these vectors corresponds to the average hydrogen bond distance, $d_\mathrm{HB}=2.8\text{ \AA}$ \cite{Jorgensen1983}, so that each vector points to the preferred position of a hydrogen bond partner. 
In addition, each bond vector is assigned a directionality, with $\bp_{1,2}$ and $\bp_{3,4}$ representing hydrogen bond donors and acceptors, respectively.
The orientation of this probe particle is specified by the vector $\vec{\kappa}=(\cos \theta_1, \cos \theta_2)$, where $\theta_1$ and $\theta_2$ specify the angles made between the donor bond vectors and the surface normal pointing away from the bulk.

The interactions of the probe molecule with the elements of the interfacial density field are governed by an empirical hydrogen bonding potential that depends on the probe molecule's position, $a$, orientation, $\vec{\kappa}$, and hydrogen bonding configuration, as specified by a set of binary variables, $\{n_k^{(\gamma)}\}$.
This potential is given by,
\begin{equation}
E(\vec{\kappa},a,\{n_k^{(\gamma)}\}) = \sum_{i = 1}^{4} \left[\epsilon_\text{w} n_i^\text{(w)}(a, \bp_i) + \epsilon_\text{s} n_i^\text{(s)}(a, \bp_i) \right]~,
 \label{eq:energy}
\end{equation}
where $\epsilon_\text{w}$ denotes the effective energy of a water-water hydrogen bond, $\epsilon_\text{s}$ represents the effective energy of a surface-water hydrogen bond, and $n_i^{(\gamma)}$ indicates the hydrogen bonding state of the $i$th bond vector to either water (\emph{i.e.}, $\gamma = \mathrm{w}$) or the surface (\emph{i.e.}, $\gamma = \mathrm{s}$). 
Specifically, $n_i^{(\gamma)} = 1$ if the probe molecule has formed a hydrogen bond of type $\gamma$ along $\bp_i$ and $n_i^{(\gamma)} = 0$ otherwise.
Here we treat each $n_i^{(\gamma)}$ as an independent random variable with statistics given by,
\begin{equation}
n_i^{(\gamma)} = \left\{  \begin{array}{l l}
    1, & \text{with probability $P_\text{HB}^{(\gamma)}(a_i)$ }~,\\
    0, & \text{with probability $1-P_\text{HB}^{(\gamma)}(a_i)$ }~,
    \end{array} \right.\
\label{eq:ni_dist}
\end{equation}
where $a_i = a - \bp_i \cdot \np$ denotes the terminal position of the $i$th bond vector, $\np$ is the unit vector normal to the plane of the interface, and $P_\text{HB}^{(\gamma)}(a_i)$ specifies the probability to form a hydrogen bond at position $a_i$ with either water (\emph{i.e.}, $\gamma = \mathrm{w}$) or surface (\emph{i.e.}, $\gamma=\mathrm{s}$).
We assume that this probability takes the simple form, $P_\text{HB}^{(\gamma)}(a_i)\propto \rho_\gamma(a_i)$, where the proportionality constant is chosen to reproduce average number of hydrogen bonds in the bulk liquid.
We also assume that these statistics are subject to a constraint that each bond vector can form only one bond (\emph{i.e.}, $\bp_i$ cannot simultaneously bond with water and surface).

The water-water hydrogen bond energy was fixed at a value of $\epsilon_\mathrm{w}=-1.77 \,k_\text{B}T$ at $T=298\mathrm{K}$, based on our previous parameterization of this model for the liquid water-vapor interface.\cite{Shin2017}
The surface-water hydrogen bond energy, $\epsilon_\mathrm{s}$, is thus a parameter that we vary in order to describe surfaces with different chemical characteristics. 
We define the properties of $\rho_\text{s}(a)$ based on the analysis of simulation data. 
Specifically, for a given model surface we computed the density of surface molecules relative to the position of the intrinsic water interface.
We then fit the leading edge of the resulting density profile to a Gaussian. 
This procedure yielded a range of means, $a_\text{s}$, and variances, $\sigma_\text{s}^2$, that ranged approximately from $a_\text{s} = -2.1 \,\text{\AA}$ to $-1.1 \,\text{\AA}$ and $\sigma_\text{s} = 0.4 \,\text{\AA}$ to $0.6 \,\text{\AA}$ for values of $\alpha = 0$ to 1, respectively. 
Specific parameters for each value of $\alpha$ are described in the SI.

In the context of this model, the probability for a molecule at position $a$ to adopt a given orientation, $\vec{\kappa}$, can thus be expressed as, 
\begin{equation}
P_\mathrm{MF}(\vec{\kappa} \vert a) = \left \langle e^{-\beta E(\vec{\kappa},a,\{n_k^{(\gamma)}\})} \right \rangle_\mathrm{b}/Z(a),
\label{eq:dist1}
\end{equation}
where $\langle \cdots \rangle_\mathrm{b}$ denotes an average over all possible hydrogen bonding states (\textit{i.e.}, variations in the $n_i^{(\gamma)}$'s), $\beta = 1/k_\text{B}T$, and $Z(a)= \int d\vec{\kappa} \left \langle e^{-\beta E(\vec{\kappa},a,\{n_k^{(\gamma)}\})} \right \rangle_\mathrm{b}$ is the orientational partition function for the probe molecule at position $a$.
By evaluating the average explicitly based on the constrained statistics of $n_i^{(\gamma)}$ as specified above, the numerator of Eq.~(\ref{eq:dist1}) can be written as,
\begin{equation}
\left \langle e^{-\beta E(\vec{\kappa},a,\{n_k^{(\gamma)}\})} \right \rangle_\mathrm{b} = \prod_{i=1}^4 \left [ 1 + P_\mathrm{HB}^\text{(w)} \left (a_i)(e^{-\beta \epsilon_\text{w}} - 1 \right ) + P_\mathrm{HB}^\text{(s)} \left (a_i)(e^{-\beta \epsilon_\text{s}} - 1 \right ) \right ].
\label{eq:dist2}
\end{equation}
Together, Eqs.~(\ref{eq:dist1}) and~(\ref{eq:dist2}) can be used to compute the orientational molecular structure of the liquid water interface.
To facilitate comparison of this mean field model to the results of atomistic simulation, we project the distribution $P_\mathrm{MF}(\vec{\kappa}\vert a)$ onto a reduced dimensional distribution,  
\begin{equation}
P_\mathrm{MF}(\cos \theta_\mathrm{OH} \vert a) = \int d\vec{\kappa} P_\mathrm{MF}(\vec{\kappa}\vert a) \left [\frac{1}{2} \sum_{i=1}^2 \delta(\cos \theta_i - \cos \theta_\mathrm{OH}) \right ],
\label{eq:red_dist}
\end{equation}
where the summation is taken over the two donor bond vectors, $\cos \theta_i = \bp_i \cdot \np / \vert \bp_i \vert$, and $\delta (x)$ is the Dirac delta function.

\subsection{Quantifying Intrinsic Hydropathy From Atomistic Simulation Data}

The characteristics of $P_\mathrm{MF}(\cos \theta_\mathrm{OH} \vert a)$ depend on the value of the surface-water hydrogen bond energy, $\epsilon_\mathrm{s}$. 
Similarly, the characteristics of $P_\mathrm{sim}(\cos \theta_\mathrm{OH} \vert a)$ computed from simulations with the molecular surfaces described in the previous section (and depicted in Fig.~\ref{fig:1} depend on the value of $\alpha$.
By comparing $P_\mathrm{MF}(\cos \theta_\mathrm{OH} \vert a)$ and $P_\mathrm{sim}(\cos \theta_\mathrm{OH} \vert a)$ we can relate values of $\alpha$ to associated values of $\epsilon_\mathrm{s}$.
This relationship thus allows us to assign a value of IH to a given surface based on atomistic simulation data.

\begin{figure}[h]
\centering
\includegraphics[width = 3.4 in]{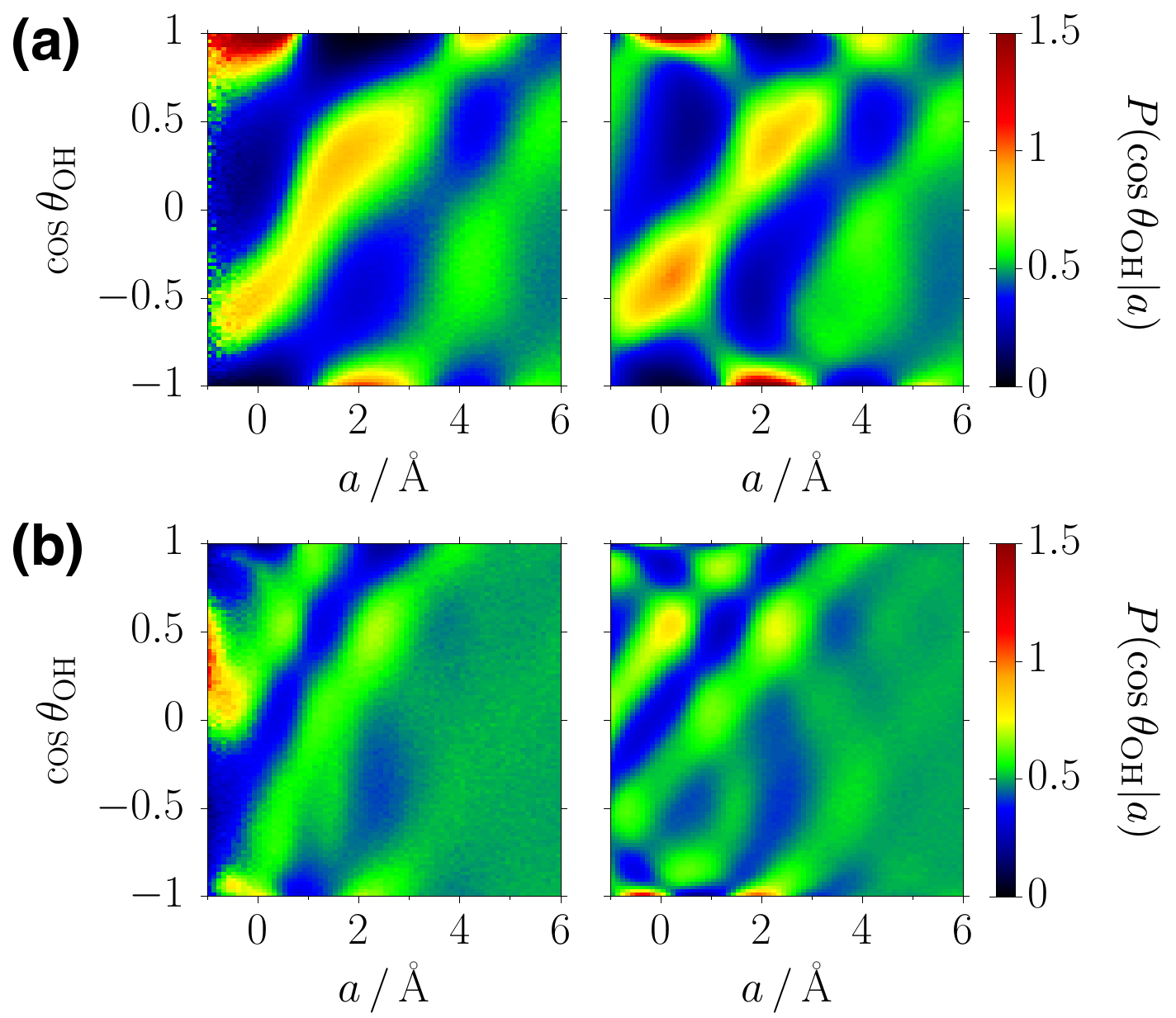}
\caption{Plots of the reduced orientational distributions for a specific surface with different polarities, (a) $\alpha = 0$ and (b) $\alpha = 1$, where color shading indicates the value of probability density. Each panel compares the result from the MD simulation (left) to that from the mean-field model (right). Model parameters used for the above plots are $\epsilon_\text{s}^* = 0 \,k_B T$ and $\epsilon_\text{s}^* = -1.55 \,k_B T$ for $\alpha = 0$ and $\alpha = 1$, respectively.}
\label{fig:6}
\end{figure}

To make a quantitative comparison between $P_\mathrm{sim}(\cos \theta_\mathrm{OH} \vert a)$ and $P_\mathrm{MF}(\cos \theta_\mathrm{OH} \vert a,\epsilon_\mathrm{s})$, where we now include the conditional dependence on $\epsilon_\mathrm{s}$ for the mean field model, we compute a fitness function $\Gamma(\epsilon_\mathrm{s})$ based on the Kullback-Leibler divergence.\cite{Kullback1951}
This fitness function is given by, 
\begin{equation}
\Gamma(\epsilon_\text{s}) = \int d a \int d (\cos\theta_\text{OH}) \, P_\mathrm{sim} (\cos\theta_\text{OH} | a)\ln \left[ \frac{ P_\mathrm{sim}(\cos\theta_\text{OH} | a)}{P_\mathrm{MF} (\cos\theta_\text{OH} | a,\epsilon_\text{s})} \right],
\label{eq:KL_div}
\end{equation}
which quantifies the similarity between the orientational molecular structure of a simulated system and that of our mean field model at a given value of $\epsilon_\mathrm{s}$. 
By minimizing $\Gamma(\epsilon_\mathrm{s})$ we can therefore identify the value of $\epsilon_\mathrm{s}$ that most closely mimics the effective surface-water interactions of the simulated system.
The value of $\epsilon_\mathrm{s}$ that minimizes $\Gamma(\epsilon_\mathrm{s})$, denoted by $\epsilon_\mathrm{s}^*$, we thus take to represent the IH of the surface.
Figure~\ref{fig:6} shows a comparison between $P_\mathrm{sim}(\cos \theta_\mathrm{OH} \vert a)$ and $P_\mathrm{MF}(\cos \theta_\mathrm{OH} \vert a,\epsilon_\mathrm{s}^*)$ for simulations with $\alpha=0$ and $\alpha=1$.
This comparison reveals that our simple model is capable of capturing the sensitivity of interfacial molecular structure to changes in surface-water interactions. 

\begin{figure}[h]
\centering
\includegraphics[width = 3.4 in]{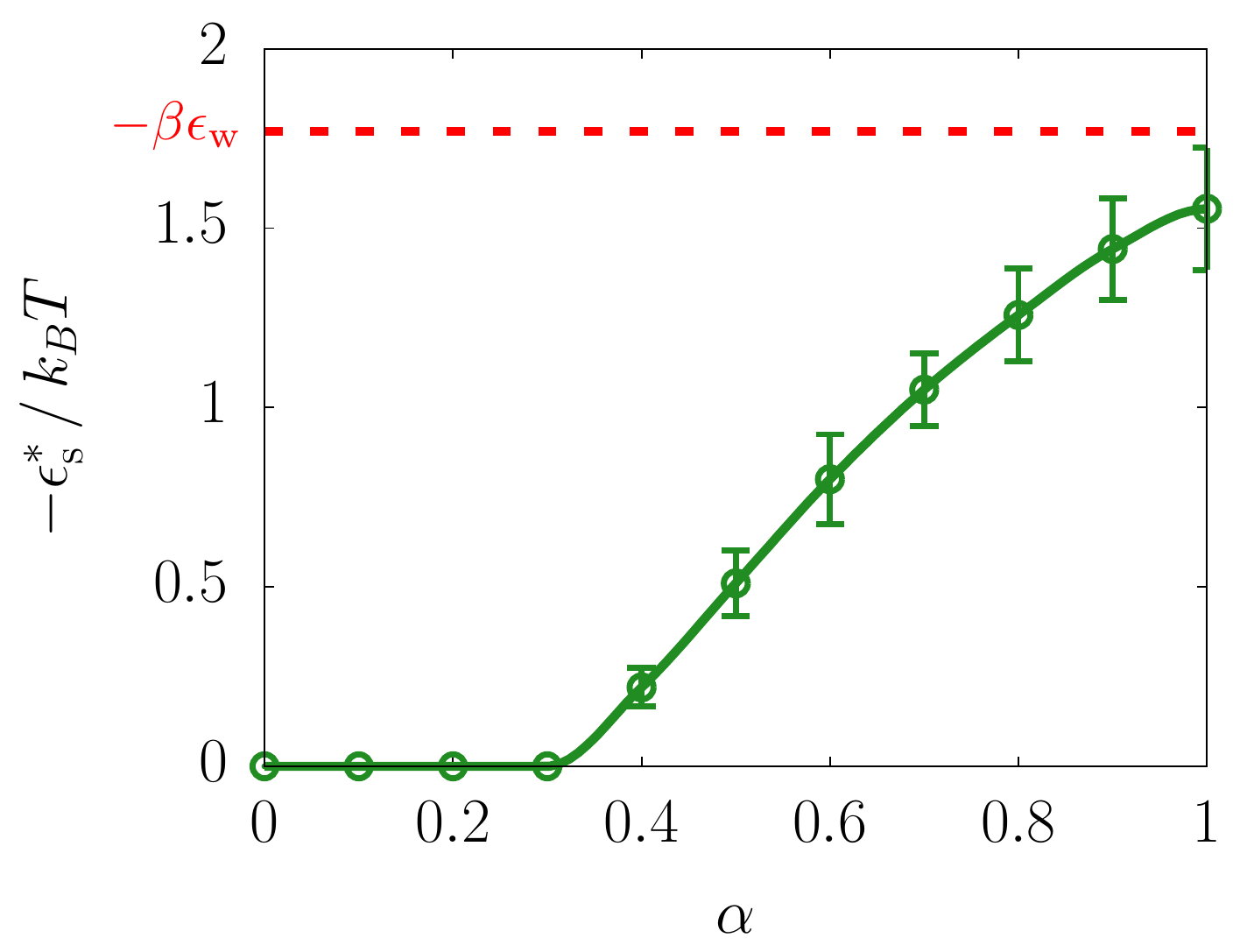}
\caption{A plot of the optimized parameter for the effective surface-water interaction against the surface polarity shown with error bars, where the green solid line is a guide to the eye and the red dashed line indicates the optimal parameter for the interaction between the liquid water molecules, $\epsilon_\text{w} = -1.77 \,k_\text{B}T$. }
\label{fig:7}
\end{figure}

In Fig.~\ref{fig:7} we plot the dependence of $\epsilon_\mathrm{s}^*$ on $\alpha$.
For $\alpha<0.4$ we observe that $\epsilon_\mathrm{s}^*\approx 0$, indicating that for these cases surface-water interactions exert a negligible influence on the structure of water's interfacial hydrogen bonding network.
For $\alpha \geq 0.4$, $\epsilon_\mathrm{s}^*$ increases monotonically with $\alpha$, reaching a value of nearly $\epsilon_\mathrm{w}$ when $\alpha=1$. 
The difference of $\epsilon_\mathrm{s}^*$ from $\epsilon_\mathrm{w}$ when $\alpha=1$ may be surprising because the surface has a force field and structure that is identical to that of bulk liquid water.
The difference between $\epsilon_\mathrm{s}^*$ and $\epsilon_\text{w}$ arises because the surface is rigid, so the effective hydrogen bonding interactions between the surface and water lack the entropic stabilization associated with hydrogen bond network flexibility.

We observe that $\epsilon_\mathrm{s}^*$ and $\langle \delta \lambda_\mathrm{phob} \rangle$ exhibit a similar dependence on $\alpha$, indicating the strong relationship between IH and variations in interfacial molecular structure. 
Notably, the behavior of $\epsilon_\mathrm{s}^*$ reveals a clear threshold that is not apparent in $\langle U_\text{w-s} \rangle$, plotted in Fig.~\ref{fig:2}.
Evidently, when $\alpha$ is small, changes in $\langle U_\text{w-s} \rangle$ with $\alpha$ do not contribute to changes in the structure of the interfacial hydrogen bonding network.
Rather they contribute to changes in the spatial profile of the intrinsic water interface, such as the mean and variance of the interfacial heights.
Thus, the properties that control aqueous interfacial solvation are determined by an interplay between the intrinsic properties of the interfacial liquid and the fluctuations in interfacial density that arise due to entropically-driven variations in the position of the intrinsic liquid interface.

By combining simulation tools for quantifying interfacial molecular structure with insight gained through a simple model of interfacial hydrogen bonding, we have highlighted that hydrophilic interfacial structure emerges through a competition between surface-water interactions and the collective water-water interactions of the bulk liquid. 
Using the mean field model, we evaluated the contribution of surface-water interactions to the emergence of hydrophilic interfacial structure in terms of the effective hydrogen bond energy, $\epsilon_\text{s}^*$.
This specific energetic component can be interpreted as a novel scale for the surface hydropathy; one that reports directly on the ability of the surface to modify water's preferred interfacial hydrogen bonding structure.
With this measure, the influence of a hydrophilic surface on a water interface can be neatly separated into its structural and spatial (\emph{e.g.} changes in capillary-wave behavior) components. 

\begin{acknowledgement}

We acknowledge useful discussions with Professors Dor Ben-Amotz, Shekhar Garde, and Amish Patel. This work was supported by the National Science Foundation under CHE-1654415 and also partially (SS) by the Kwanjeong Educational Foundation in Korea.

\end{acknowledgement}

\providecommand{\latin}[1]{#1}
\providecommand*\mcitethebibliography{\thebibliography}
\csname @ifundefined\endcsname{endmcitethebibliography}
  {\let\endmcitethebibliography\endthebibliography}{}

\clearpage
\setcounter{equation}{0}
\setcounter{figure}{0}
\setcounter{table}{0}
\setcounter{page}{1}
\makeatletter
\renewcommand{\theequation}{S\arabic{equation}}
\renewcommand{\thefigure}{S\arabic{figure}}
\renewcommand{\bibnumfmt}[1]{[S#1]}
\renewcommand{\citenumfont}[1]{S#1}

\section*{Supporting Information}

\subsection*{Simulation setup}
Here we provide the details of simulation setup. 
Molecular geometries of the non-surface water molecules were constrained with the SHAKE algorithm.
Particle Mesh Ewald was used to handle the long-range part of electrostatic interactions with the periodic boundary conditions in all directions. 
Cutoff distance for the long-range interaction was 10 \AA.
Propagation of dynamics was based on the standard velocity-Verlet integrator with a time step of 2 fs. 
The population of liquid water molecules were coupled to the Langevin thermal bath at $T=298 \,\text{K}$ every 0.1 ps. 

\subsection*{Interfacial molecular structure versus excess chemical potential} 
In order to relate the aqueous interfacial molecular structure to the surface hydropathy, we compare values of $\delta \lambda_\text{phob}$ to quantity, $\Delta\mu_\text{ex} = \mu_\text{ex}^{(\text{int})} - \mu_\text{ex}^{(\text{bulk})}$, where $\mu_\text{ex}^{(\text{int})}$ and $\mu_\text{ex}^{(\text{bulk})}$ are the excess chemical potentials for inserting a hard-sphere solute near the interface and in the bulk water, respectively. For a given location over a hydrated surface, namely $\rp_\text{surf}$, the excess chemical potential near the interface can be specified as 
\begin{equation}
\mu_\text{ex}^{(\text{int})} (\rp_\text{surf}) = -k_B T \ln P_v(0|\rp_\text{surf})~,
\end{equation}
where $P_v(0|\rp_\text{surf})$ is the probability that no solvent molecule is observed within a spherical cavity of radius $R$ centered at $\rp_\text{surf} + R\hat{z}$ (\emph{i.e.}, the cavity contacts with the surface at $\rp_\text{surf}$). This probability is computed from the simulation as
\begin{equation}
P_v (0|\rp_\text{surf}) = \left\langle \delta \left( \sum_i H( R - |\rp_i - \rp_\text{surf} - R \hat{z}| ) \right )\right\rangle~,
\end{equation}
where the angle bracket denotes an equilibrium average, $\delta(x)$ is a Dirac-delta function, $H(x)$ is a Heaviside step function, and the summation is taken over all liquid water molecules for their positions, $\rp_i$. 
For $\rp_\text{surf}$, the surface boundary was determined from a Willard-Chandler interface \cite{Willard:2010da} constructed for the molecules belonging to the solid phase. 
Similarly, the probability of a cavitation was computed in the bulk, which gave $\mu_\text{ex}^{(\text{bulk})} = 4.0 \,k_\text{B} T$ for $R = 2.5 \text{ \AA}$. 

\begin{figure}[h!]
\centering
\includegraphics[width = 6.4 in]{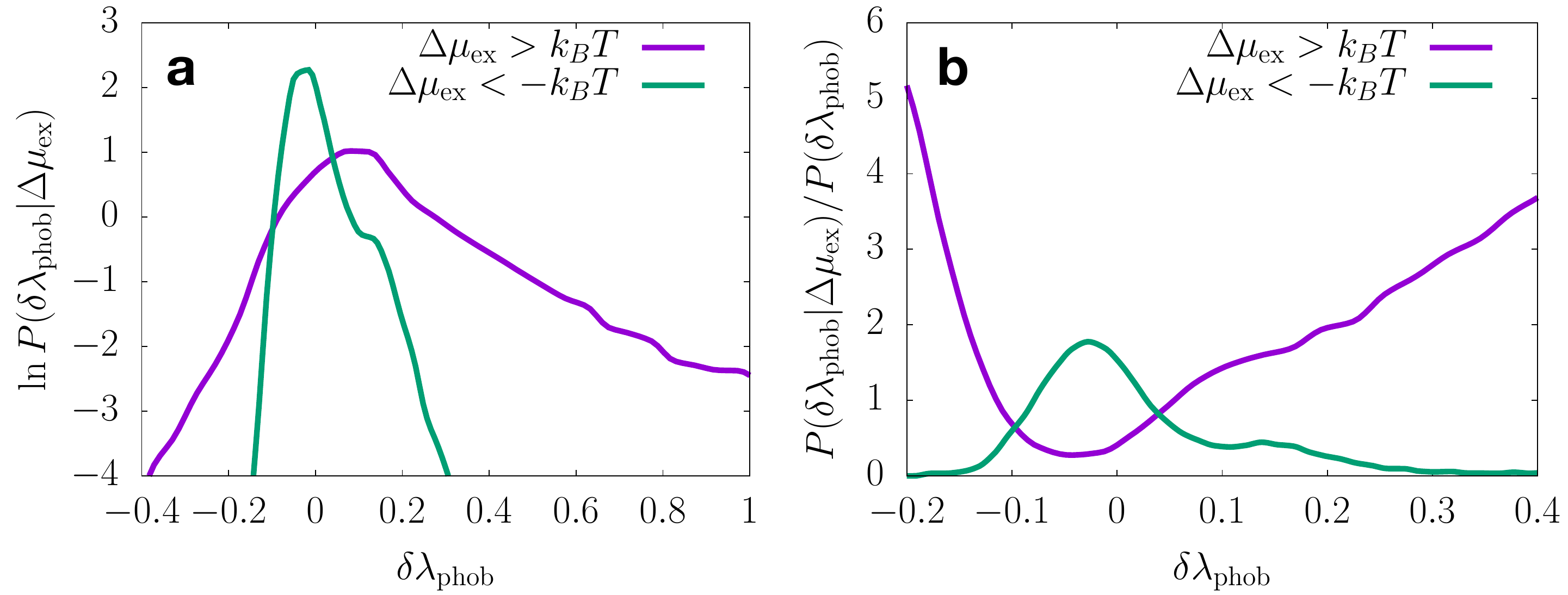}
\caption{(a) Plots of conditional probability distributions of $\delta \lambda_\text{phob}$ given $\Delta\mu_\text{ex} > k_\text{B}T$ and $\Delta\mu_\text{ex} < -k_\text{B}T$. (b) Plots of the conditional probability distributions normalized by $P(\delta \lambda_\text{phob})$. Although the panel b shows that the righthand crossing point between the curves is $\delta \lambda_\text{phob} \approx 0.5$, we set the upper bound of hydrophobic interfacial molecular structure to be $\delta \lambda_\text{phob} = 1.0$ where $P(\delta \lambda_\text{phob}|\Delta\mu_\text{ex} > k_\text{B}T)$ reaches the maximum and $P(\delta \lambda_\text{phob}|\Delta\mu_\text{ex} < -k_\text{B}T)$ is featured with some shoulder as shown in panel a.}
\label{fig:S1}
\end{figure}

Based on the values of $\delta \lambda_\text{phob}(\rp_\text{surf})$ and $\Delta\mu_\text{ex}(\rp_\text{surf})$ for all surface positions, we computed conditional probability distributions, $P(\delta \lambda_\text{phob}|\Delta\mu_\text{ex})$. 
Here, we identify the positions of the surface with $\Delta\mu_\text{ex} < -k_\text{B}T$ as being hydrophobic and those with $\Delta\mu_\text{ex} > k_\text{B}T$ as being hydrophilic. 
As illustrated in Fig.~\ref{fig:S1}, $P(\delta \lambda_\text{phob}|\Delta\mu_\text{ex} < -k_\text{B}T)$ is dominant over $P(\delta \lambda_\text{phob}|\Delta\mu_\text{ex} > k_\text{B}T)$ in the range of  $\left\vert \delta \lambda_\text{phob} \right\vert \lesssim 0.1$, and thus we interpret the values of $-0.1 \le \delta \lambda_\text{phob} \le 0.1$ as hydrophobic interfacial molecular structure and values of $|\delta \lambda_\text{phob}| > 0.1$ as hydrophilic interfacial molecular structure. 

\subsection*{Density profiles for model surfaces} 
\begin{figure}[h!]
\centering
\includegraphics[width = 6.4 in]{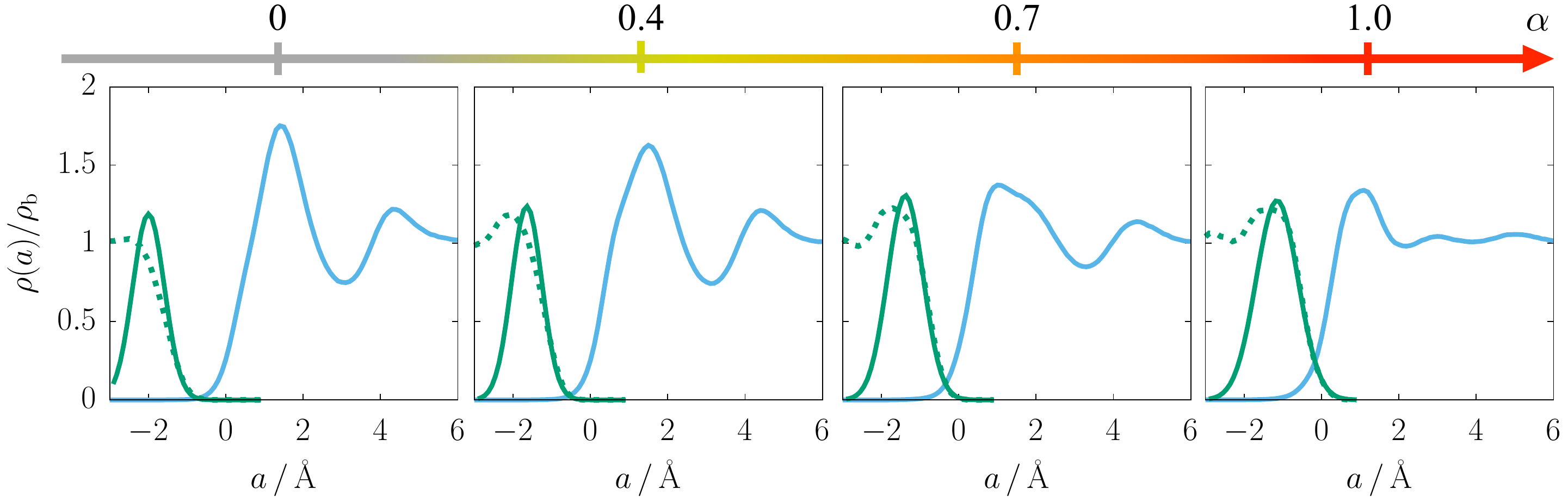}
\caption{Plots of density profiles for different values of surface polarity. $\rho_\text{w}(a)$ is rendered by blue line. $\rho_\text{s}^\text{(int)}(a)$ and $\rho_\text{s}(a)$ are rendered by green dotted and solid lines, respectively.}
\label{fig:S2}
\end{figure}

As described in the main text, we computed the density profiles of the molecules in liquid and surface, namely $\rho_\text{w}(a)$ and $\rho_\text{s}^\text{(int)}(a)$, relative to the position of the intrinsic water interface, following the procedure described in Ref. \citenum{Willard:2014js}. 
These density profiles change significantly as the surface polarity, $\alpha$, increases from 0 to 1. 
As illustrated in Fig.~\ref{fig:S2}, the density profile for the surface gets closer to that for the liquid water upon the increase in $\alpha$, which indicates more adsorption of the solvent molecules to the surface of larger polarity. 
For the mean-field model, we use a Gaussian function as an effective surface density profile, which is fitted from the leading peak of $\rho_\text{s}^\text{(int)}(a)$ (more specifically, the region of $a \ge \argmax_a\{\rho_\text{s}^\text{(int)}(a)\}$).
The Gaussian-fitted density profile is parametrized as $\rho_\text{s}(a)/\rho_\text{b} = \rho_0 e^{-(a - a_\text{s})^2/2\sigma_\text{s}^2}$, where $\rho_\text{b}$ is the bulk density of liquid water, such that the set of parameters, $(a_\text{s}, \sigma_\text{s}, \rho_0)$, well represents the mean characteristics of the first molecular layer of a given surface. 
Table~\ref{tab:1} lists a set of parameters for a specific model surface of the polarity ranging from $\alpha = 0$ to 1.
Note that the Gaussian mean and width, $a_\text{s}$ and $\sigma_\text{s}$, show the trends of increasing along with $\alpha$.

\begin{table}[h!]
\caption{\label{tab:1}%
Parameters for the Gaussian-fitted density profile, $\rho_\text{s}(a)$, for a specific model surface with varied polarity.
}
\centering
\begin{tabular}{ c | r  r  c}
\hline\hline
$\alpha$ & $a_\text{s} \,\text{(\AA)}$ & $\sigma_\text{s} \,\text{(\AA)}$ & $\rho_0$ \\
\hline
0  &  -1.99762  &  0.411275  &  1.18779  \\
0.1  &  -2.07123  &  0.428661  &  1.47521  \\
0.2  &  -1.94063  &  0.428272  &  1.2517  \\
0.3  &  -1.82815  &  0.426502  &  1.2016  \\
0.4  &  -1.64603  &  0.394326  &  1.23684  \\
0.5  &  -1.56485  &  0.410636  &  1.28688  \\
0.6  &  -1.52418  &  0.449296  &  1.36765  \\
0.7  &  -1.38108  &  0.458931  &  1.30655  \\
0.8  &  -1.34561  &  0.508773  &  1.33211  \\
0.9  &  -1.24132  &  0.527848  &  1.282  \\
1  &  -1.14727  &  0.539936  &  1.27362  \\
\hline\hline
\end{tabular}
\end{table}

\end{document}